\documentclass[twocolumn,journal]{IEEEtran}

\usepackage[T1]{fontenc}
\usepackage[cmex10]{amsmath}
\usepackage{amssymb}
\usepackage{amsthm}
\usepackage{acronym}
\usepackage{threeparttable}
\usepackage{booktabs}
\usepackage{enumitem}
\usepackage[pdftex]{graphicx}
\usepackage{cite}
\usepackage{color}
\usepackage{soul}

\usepackage{amsmath}
\DeclareMathOperator*{\argmin}{arg\,min}
\DeclareMathOperator*{\argmax}{arg\,max}
\usepackage{comment}

\usepackage{tabularx}  
\usepackage{algorithm}
\usepackage{algpseudocode}
\newtheorem*{theorem*}{Theorem}
\newtheorem{theorem}{Theorem}
\usepackage{subcaption}
\usepackage{float}
\usepackage[font=small]{caption}
\usepackage[unicode=true,
 bookmarks=true,bookmarksnumbered=true,bookmarksopen=true,bookmarksopenlevel=1,
 breaklinks=false,pdfborder={0 0 0},pdfborderstyle={},backref=false,colorlinks=false]
 {hyperref}
\hypersetup{pdftitle={Your Title},
 pdfauthor={Your Name},
 pdfpagelayout=OneColumn, pdfnewwindow=true, pdfstartview=XYZ, plainpages=false}

\AtBeginDocument{%
  \setlength{\abovedisplayskip}{4pt}
  \setlength{\belowdisplayskip}{4pt}
  \setlength{\abovedisplayshortskip}{4pt}
  \setlength{\belowdisplayshortskip}{4pt}
}
\DeclareMathAlphabet{\mathpzc}{OT1}{pzc}{m}{it}

\newcommand{\bfg}[1]{\boldsymbol{#1}}

\acrodef{rl}[RL]{rate limiter}
\acrodef{tbcd}[TBCD]{twin-based control defense}
\acrodef{iss}[ISS]{input-to-state-stable}
\acrodef{der}[DER]{distributed energy resource}
\acrodef{dg}[DG]{distributed generation}
\acrodef{dso}[DSO]{distribution system operator}
\acrodef{ev}[EV]{electric vehicle}
\acrodef{pcc}[PCC]{point of common coupling}
\acrodef{pll}[PLL]{phase-locked loop}
\acrodef{qep}[QEP]{quadratic eigenvalue problem}
\acrodef{srf}[SRF]{synchronously rotating frame}
\acrodef{tso}[TSO]{transmission system operator}
\acrodef{vsi}[VSI]{voltage source inverter}
\acrodef{res}[RES]{renewable energy source}
\acrodef{ess}[ESS]{energy storage system}
\acrodef{lec}[LEC]{local energy community}
\acrodef{lem}[LEM]{local energy market}
\acrodef{dgu}[DGU]{distributed generation unit}
\acrodef{cpl}[CPL]{constant power load}
\acrodef{gfl}[GFL]{grid-following}
\acrodef{gfm}[GFM]{grid-forming}
\acrodef{ibr}[IBR]{inverter-based resource}
\acrodef{rsu}[RSU]{robust synchronization unit}
\acrodef{iot}[IoT]{Internet-of-Things}
\acrodef{scada}[SCADA]{supervisory control and data acquisition}
\acrodef{pmu}[PMU]{power measurement unit}
\acrodef{hmi}[HMI]{human machine interface}
\acrodef{fft}[FFT]{fast fourier transform}
\acrodef{chil}[CHIL]{controller hardware-in-the-loop}
\acrodef{fdi}[FDI]{false data injection}
\acrodef{pe}[PE]{persistently exciting}
\acrodef{mrac}[MRAC]{model reference adaptive control}
\acrodef{ugas}[UGAS]{uniformly globally asymptotically stable}
\acrodef{uges}[UGES]{uniformly globally exponentially stable}
\acrodef{lmi}[LMI]{linear matrix inequality}
\acrodef{dsp}[DSP]{digital signal processor}
\acrodef{facts}[FACTS]{flexible AC transmission system}
\acrodef{dc}[DC]{data center}
\acrodef{llm}[LLM]{large language model}
\acrodef{psu}[PSU]{power supply unit}
\acrodef{ups}[UPS]{uninterruptible power supply}
\acrodef{ai}[AI]{artificial intelligence}
\acrodef{laa}[LAA]{load altering attack}
\acrodef{ltp}[LTP]{linear time periodic}
\acrodef{phil}[PHIL]{power hardware-in-the-loop}
\makeatletter
\let\oldforeign@language\foreign@language
\DeclareRobustCommand{\foreign@language}[1]{%
  \lowercase{\oldforeign@language{#1}}}
\theoremstyle{plain}

\theoremstyle{remark}

\makeatother

\definecolor{blue1}{RGB}{33,92,175}

\title{Rented Oscillations: Synchronization Attacks on the Grid Through Data Center Workloads}
\title{\huge{Rented Oscillations: Multi-Tenant Data Center Workload Synchronization Attacks and Mitigation}}

\author{Zaint A. Alexakis,  Li Zhou, S. Gokul Krishnan, and Charalambos Konstantinou}

\begin{document}
\IEEEaftertitletext{\vspace{-2.5\baselineskip}}
\maketitle
\pagestyle{plain} 
\IEEEpeerreviewmaketitle
\bstctlcite{IEEEexample:BSTcontrol}

\begin{abstract}
Multi-tenant \ac{dc} facilities rent computing capacity on demand, placing controllable load within reach of anyone willing to pay for it. This paper considers a stealthy adversary who rents workstations and runs deliberately unoptimized \ac{llm} training, synchronizing their consumption into large periodic oscillations while each workload remains individually indistinguishable from benign training. The periodicity of such workloads is characterized experimentally and used to model facility consumption, showing that oscillations emerge from the interaction of concurrent workloads, whether aligned by an adversary or arising naturally. A combined small- and large-signal assessment, under a hardware-grounded threat model whose feasibility is supported by experiments on a supercomputer, locates the oscillatory weak points of the network. MW-scale oscillations are shown to saturate the generators' control limiters, amplifying their effect, while Floquet analysis reveals a lightly damped mode emerging near the forcing frequency. To counter both origins, signal-energy-based detection and mitigation are proposed, the latter re-timing the workloads to minimize the facility's oscillatory energy. The approach is validated through \ac{phil} experiments and real-time simulations on a surrogate model of the Saudi Arabian transmission network. 
\end{abstract}
\acresetall
\vspace{-2mm}
\IEEEpeerreviewmaketitle{}
\setlength{\textfloatsep}{0pt}
\vspace{-7mm}
\section{Introduction}
\vspace{-1mm}
The rapid adoption of \ac{ai} and cloud services has turned \acp{dc} into one of the fastest-growing loads on the power grid, introducing operational challenges and new attack surfaces. In the United States alone, serving them is projected to require 148~GW of interconnection capacity by 2030~\cite{lbnl2026datacenter}. This growth spans facilities of very different nature, from frontier sites dedicated to training a single massive model to multi-tenant facilities that rent their workstations to the public. As these loads grow, so does their footprint on grid operation, and \ac{dc}-related incidents have already been reported across North America~\cite{nerc2025largeloads}, such as the sustained oscillations originating from a \ac{dc} in the Dominion Energy system~\cite{147MISHRA2025101735}. Scrutiny, however, has concentrated on frontier-scale facilities, whose volatility is conspicuous, while the far more numerous multi-tenant facilities have drawn little attention, owing to their presumably steadier demand.

Behind this presumption lies the diversity of the workloads such facilities host. A multi-tenant facility runs hundreds of powerful workstations, whose jobs range from power-hungry \ac{ai} training to inference and, more recently, agentic tasks built on the trained models \cite{agentic,infere}. Training cycles through computation and communication phases, drawing a periodic profile that alternates between high and low power, whereas inference produces random bursts driven by user queries, and CPU-intensive tasks draw near-constant power \cite{hao_paper,jimenezruiz2026datacentre}. Being mutually uncorrelated, these workloads largely average out, and rental quotas, which cap the share of any single tenant, flatten the aggregate further. This steadiness, however, is not guaranteed: when training workloads happen to converge in frequency, their oscillations no longer cancel but reinforce, and the facility may oscillate even under normal operation, undermining the very assumption that it is benign. Such synchronization has been predicted theoretically to arise through the coupling introduced by facility power caps~\cite{competition_tardif2026colocatedaitrainingjobs}, without, however, examining how the characteristics of the workloads themselves govern it. More concerning still, the same mechanism can be triggered deliberately, turning these facilities into an ideal instrument for an adversary seeking to reshape their demand against the very grid that powers them.


Tenants have recently been shown to inject kHz-range power modulations into DER-rich local systems~\cite{ji2026bit2wattcyberphysicalvulnerabilityexploiting}, although such content is largely absorbed by the power electronic interface of the facility, limiting its reach beyond the local system. The adversary considered here instead exploits the sub-10~Hz band identified in recent studies~\cite{hao_paper}, which passes through this interface and targets the electromechanical dynamics of the transmission network. Rental quotas offer little protection against such an adversary: by exploiting payment fraud (e.g., stolen credit-card credentials) and identity fraud (e.g., forged identities), an attacker can acquire and run many of these high-power workstations in lockstep with deliberately inefficient training jobs, whose deepened periodic swings combine into a large oscillating load that these facilities are neither designed nor equipped to withstand. Nor is the adversary confined to a single facility, since workstations rented across a region can collectively appear, from the grid's perspective, as a single aggregate load. In contrast, existing load-altering attack vectors~\cite{insurance_11271785} must first compromise vast populations of distributed loads, such as air-conditioning units. Here, control over a large aggregate load is effectively purchased, at a cost small relative to the potential harm it can inflict.

The impact of such an attack can be assessed by examining how the oscillations propagate through the network and interact with regional assets, eroding stability margins, degrading power quality, and, where protective relays operate, interrupting supply~\cite{oscila_review_10536656}. A substantial body of work analyzes such oscillations through modal tools, locating their origin and tracing their propagation across the network~\cite{hybrid_9712201,find_7312503}, including studies that account for load dynamics and their effect on electromechanical modes~\cite{load_osc8395034}. These approaches, however, take the forcing as given rather than examining how the load generates it. Closest to the present setting,~\cite{hao_paper} models \ac{dc}-induced oscillations and their effect on the small-signal behavior of the grid, yet it concentrates on resonant modes, as do prior forced oscillation analyses~\cite{patent_7041242,forced_8982201}, and overlooks how the interaction of the workloads shapes the forcing. What remains missing is a combined small- and large-signal assessment built on a hardware-grounded \ac{dc} workload model, one that captures the interaction of the load with the multiple modes of the system and treats the workload's periodicity not as a static input but as a forcing that reshapes stability along the periodic trajectory.


On the mitigation side, oscillations from frontier training sites, the natural counterpart of this threat, are mitigated on the grid side, through devices such as hybrid storage~\cite{mit_11693182} and unified power flow controllers~\cite{upfc4745822}, or through tailored control schemes of regional assets~\cite{con_mit2_9094043}. These solutions, however, treat the symptom rather than its cause, and demand substantial capital investment or the redispatch of regional assets to accommodate the behavior of the \ac{dc}, which asset owners have little incentive to provide. A more practical direction is taken in~\cite{gov_8123888}, where the sustained oscillation is suppressed by retuning the governor parameters that shape it, though the achievable improvement is limited. Source-side remedies have also begun to emerge in industry: to suppress the periodicity of the workload,~\cite{choukse2025powerstabilizationaitraining} modulates the commanded power waveform, but at the cost of substantial wasted energy. There is therefore a need for a systematic mitigation tailored to \acp{dc}, one that exploits their intrinsic characteristics to suppress the oscillation economically at its source rather than burdening uninvolved regional assets, and that applies equally to adversarial and benign workload execution.

To address these gaps, this paper examines how multi-tenant \acp{dc} can impact the grid, and how a stealthy adversary, renting their workstations under forged identities, can deliberately exploit them. The threat model is grounded in measured workload power profiles and demonstrated on a top-20 supercomputer. Building on these measurements, a facility model is developed that reveals, under normal operation, the steadiness of such facilities to be not inherent but contingent on how their workloads interact. On this basis, a vulnerability and impact assessment combining small- and large-signal analyses identifies network locations most exposed to such oscillations, and shows that, once MW-scale, the forcing can saturate the control limiters of generators, amplifying its own impact and giving rise to lightly damped modes, as corroborated by Floquet analysis. Finally, as a foundation for policies governing workload execution, a signal-energy-based detection scheme attributes irregularities in facility demand to responsible workloads, while a mitigation algorithm turns the mechanism the adversary exploits into a safeguard, aligning and interleaving the workloads so that their oscillations cancel at a cost in throughput, and extending to geographically distributed facilities. Both mechanisms are validated through \ac{phil} experiments and real-time simulations on a surrogate model of the Saudi Arabian transmission network, demonstrating that the same workloads capable of shaking the grid can, once their execution is coordinated, coexist with it.

\vspace{-3mm}
\section{Threat Model}
\label{sec:threat}
Multi-tenant \acp{dc} place computing capacity, and hence controllable power, within reach of anyone willing to pay for it. Rental quotas limit the share of the facility that a single tenant can occupy, preventing \ac{llm} training at a scale the facility is not designed to sustain, yet they can be circumvented through multiple accounts under forged identities. This work considers a well-resourced adversary who exploits this gap to rent a significant share of the facility's GPUs from rental services, to covertly synchronize their consumption into a large oscillating load against the grid. The adversary rents full GPUs/CPUs, i.e., whole physical devices rather than fractional instances, and thus holds exclusive control over each device throughout the lease, launching deliberately unoptimized training tasks in a coordinated manner. All devices are rented within the same region, which rental platforms permit, so that the resulting oscillation is concentrated on the same portion of the grid.

\begin{figure}[t]
  \centering\includegraphics[width=0.75\columnwidth]{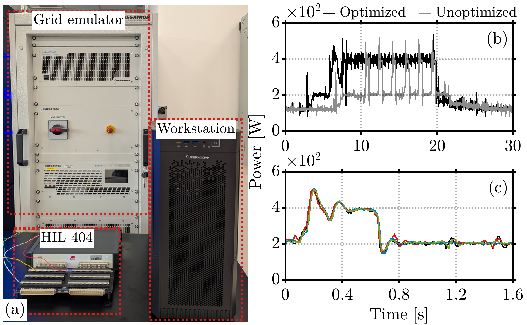}
  \vspace{-2mm}
  \caption{Testbench (a) and hardware-measured power traces of optimized and unoptimized \ac{llm} training workloads (b), with the iterations of the latter superimposed (c).}
  \label{fig:llm}
  \vspace{-1mm}
\end{figure}
What makes such an attack both feasible and covert is the inherent periodicity of unoptimized \ac{llm} training. This is demonstrated in Fig.~\ref{fig:llm}, which illustrates the power drawn by an optimized and an unoptimized training workload executed on an RTX~PRO~4500~Blackwell. While the optimized workload maintains a nearly constant consumption, the unoptimized one alternates between compute and idle phases, forming an approximately rectangular wave whose iterations, when superimposed, virtually coincide. Moreover, each return to the compute phase is accompanied by an overshoot and ringdown, attributed to the response of the \ac{psu}. When synchronized across many devices, such workloads therefore become a Trojan horse: each is individually indistinguishable from legitimate, albeit inefficient, training, yet together they form a large oscillation. The same oscillation may equally arise from undeclared training or from benign workloads that gradually converge to common frequency bands. Consequently, since the effect remains identical regardless of its cause, the computing side alone can hardly distinguish between the two. The power traces, however, inevitably reveal how the workloads interact, and it is from this power-system perspective that the threat is addressed in this work.

\begin{figure}[t]
  \centering
  \includegraphics[width=0.8\columnwidth]{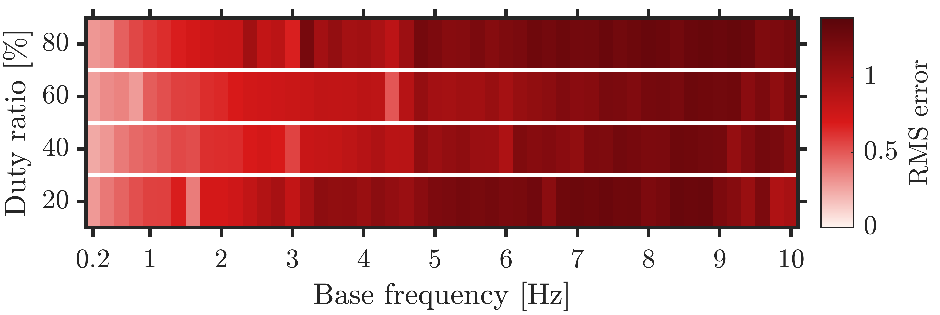}
  \vspace{-2mm}
  \caption{Normalized RMS error between the commanded and measured GPU power as a function of duty ratio and base frequency.}
  \label{fig:heat}
  \vspace{-3mm}
\end{figure}

While the adversary's objectives are clear, workload controllability remains uncertain and shapes the attack's vulnerability assessment, detection, and mitigation. Unlike most cyber-attack studies, this work derives attacker capabilities from the hardware available to the adversary rather than assuming them. To this end, the testbench of Fig.~\ref{fig:llm} is programmed to drive the GPU with duty ratios of 20-80\,\% and base frequencies of 0.2-10\,Hz, ten cycles each, while the power drawn from its PSU is recorded. Each block is then scored by the normalized RMS error between the measured power and an ideal rectangle of the same frequency, duty ratio, and energy, as shown in Fig.~\ref{fig:heat}. The GPU reproduces the rectangle most closely below about 2\,Hz, degrades progressively up to about 5\,Hz, and beyond that no longer follows the commanded period. The capability of the adversary is therefore bounded at about 5\,Hz, with controllability highest at intermediate duty ratios (40-60\,\%) and lowest at the extremes. These limits delimit frequencies that local \ac{llm} training can exhibit, and since distributed execution can only narrow this band due to synchronization overhead, they are adopted as the bounds of the threat model and define the focus of the analysis.

\vspace{-4mm}
\subsection{Feasibility Assessment}
\begin{figure}[t]
  \centering
  \vspace{-2mm}\includegraphics[width=0.8\columnwidth]{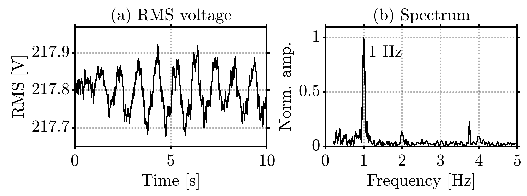}
  \vspace{-3mm}
  \caption{Emulated workload synchronization attack on a supercomputer, observed at a laboratory sharing the same substation.}
  \label{fig:super_attack}
  \vspace{-1mm}
\end{figure}

\vspace{-1mm}
To demonstrate the feasibility of the threat model, an up-scaled experiment was conducted, with institutional authorization, on one of the world's top-20 supercomputers. Since GPU partitions were not yet available at the time of the study, several allocated CPU nodes were coordinated to modulate their load as a rectangular wave with a 50\% duty ratio and a fundamental frequency of 1~Hz. As the supply voltage of the facility is not externally accessible, the effect was observed at the authors' laboratory, located approximately 1~km away and connected to the same network through a shared substation. The results are depicted in Fig.~\ref{fig:super_attack}, which shows the RMS voltage of a single phase: once the attack is launched at $t\approx1$~s, the voltage begins to oscillate at 1~Hz, as confirmed by its \ac{fft}. The resulting peak-to-peak oscillation of 0.2~V is modest, reflecting the small share of computing power acquired and the energy-storage buffers, such as batteries and flywheels, reported to support the machine; it nevertheless shows that coordinated workloads propagate through the network and are observable beyond the facility itself.

\vspace{-3mm}
\section{Multi-Tenant \ac{dc} Modeling}
\vspace{-1mm}
This section models the power profile of a multi-tenant \ac{dc} using measured \ac{llm} training traces from Sect.~II. Individual workloads are modeled first, then combined to characterize how their interactions shape facility demand and inform mitigation. The model also captures the power electronic interface between the GPUs and grid. The resulting profile underpins the vulnerability and impact assessment for both adversarial and benign operation.



\vspace{-3mm}
\subsection{AI-Training Workloads}
\vspace{-1mm}
Based on the waveform in Fig.~\ref{fig:llm}, each training workload is modeled as a pulse train alternating between a high-power compute (up) phase and a low-power idle (down) phase, in which GPUs wait for data over PCIe or InfiniBand during communication, as assumed in \cite{jimenezruiz2026datacentre,krajacic2026forced,hao_paper}. The $j$-th workload is characterized by its iteration rate $f_{0,j}\sim\mathcal{U}(\underline{f}_0,\overline{f}_0)$, duty ratio $r_j\sim\mathcal{U}(\underline{r},\overline{r})$, nominal compute-phase power $\hat{p}_j$, with $\ln\hat{p}_j\sim\mathcal{N}(\mu_p,\sigma_p^2)$, and initial phase $\varphi_j\sim\mathcal{U}(0,2\pi)$, all drawn once and held for the whole run. Its $i$-th iteration starts at $t_{j,i}=(i-1-\varphi_j/2\pi)/f_{0,j}$, and it draws $p_j(t)=\hat{p}_j\big(1+\Delta_{j,i}^{\mathrm{up}}+\eta_j(t)\big)$ for the first $r_j/f_{0,j}$ of it and $\hat{p}_j\big(1-\Delta_{j,i}^{\mathrm{down}}+\eta_j(t)\big)$ for the remaining $(1-r_j)/f_{0,j}$. The deviations $\Delta_{j,i}^{\mathrm{up}}\sim\mathcal{N}(0,\sigma_{\Delta}^2)$ and $\Delta_{j,i}^{\mathrm{down}}\sim\mathcal{N}(\mu_{\Delta},\sigma_{\Delta}^2)$, with $\mu_{\Delta}$ the nominal relative power reduction in the idle phase, vary the power levels between iterations, while $\eta_j(t)$, drawn at each time step from $\mathcal{N}(0,\sigma_{\eta,\mathrm{up}}^2)$ or $\mathcal{N}(0,\sigma_{\eta,\mathrm{down}}^2)$ in the compute or idle phase, respectively, captures fast intra-phase fluctuations. With $N_w$ independently executed training workloads, the aggregate AI-training demand of the \ac{dc} is $P^{\mathrm{AI}}(t)=\sum_{j=1}^{N_w}p_j(t)$.

Beyond training, such facilities host a substantial share of inference and, increasingly, agentic workloads, which are far less compute-intensive and typically occupy a single GPU or only a few. These workloads mostly serve \acp{llm} and are non-periodic, alternating between an idle phase, in which the model awaits a request, and a high-power phase, triggered when a user submits a query \cite{agentic,infere}. Since the queries arrive independently, their aggregate forms a sum of random power impulses that, by the law of large numbers, converges to a nearly constant profile. CPU-intensive workloads, which draw near-constant power while active, contribute similarly, and both are therefore represented in this work as a constant component of the facility demand.

\vspace{-3mm}
\subsection{Power Electronic Interface Dynamics}
\vspace{-1mm}
The computing infrastructure is interfaced to the grid through power electronics which, in small-signal terms, act as a filter on the workload profile. In a typical configuration~\cite{model11576565}, a back-to-back \ac{ups} converter decouples the facility from the grid, while the \ac{psu} of each server rectifies the supply through a power factor correction stage and steps it down to the GPU voltage through a dc/dc stage, omitted here as its kHz bandwidth lies well above the range of interest. Representing the \acp{psu} by an aggregate equivalent and neglecting line losses, the dc-link dynamics of the \ac{psu} and \ac{ups} are expressed through their stored energy $y_x=0.5\,c_xv_x^2$, $x\in\{\mathrm{p},\mathrm{u}\}$, which renders them linear:
\begin{subequations}
\begin{align}
  y_{\rm p}^{\prime}&=P_{\rm p}-P^{\rm AI}, & P_{\rm p}&=k_{p,\rm p}e_{\rm p}+k_{i,\rm p}\textstyle\int e_{\rm p}\,dt,\\
  y_{\rm u}^{\prime}&=P_{\rm g}-P_{\rm p}, & P_{\rm g}^{\star}&=k_{p,\rm u}e_{\rm u}+k_{i,\rm u}\textstyle\int e_{\rm u}\,dt,\\
  T_{\rm c}P_{\rm g}^{\prime}&=P_{\rm g}^{\star}-P_{\rm g}, & &
\end{align}
\end{subequations}
where $c_x$ and $v_x$ are the dc-link capacitance and voltage, $e_x=y_x^{\star}-y_x$ is the deviation from the reference energy $y_x^{\star}$ corresponding to the nominal voltage, $P_{\rm p}$ is the power drawn by the power factor correction stage, $P_{\rm g}$ is the grid-side power, which tracks its reference through an inner current loop with time constant $T_{\rm c}$, and $k_{p,x}$, $k_{i,x}$ are the PI gains. The workload profile thus reaches the grid through:
\begin{equation}\small
  \frac{P_{\rm g}(s)}{P^{\rm AI}(s)}=\frac{k_{p,\rm p}s+k_{i,\rm p}}{s^2+k_{p,\rm p}s+k_{i,\rm p}}\cdot\frac{k_{p,\rm u}s+k_{i,\rm u}}{T_{\rm c}s^{3}+s^2+k_{p,\rm u}s+k_{i,\rm u}},
\end{equation}
whose gains follow industrial practice for dc-link regulation, which limits the controller bandwidth to balance harmonic distortion against performance.
\vspace{-3mm}
\subsection{Grid-Interfacing \ac{dc} Profile}
\begin{figure}[t]
    \centering
    \vspace{-2mm}
    \includegraphics[width=0.8\linewidth]{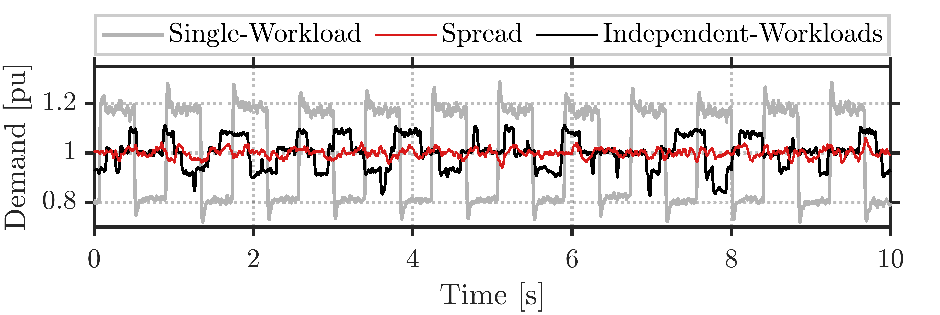}
    \vspace{-2mm}
\caption{Facility power profiles under different workload compositions.}
\label{fig:pow_profile}
\end{figure}

\vspace{-1mm}
The aggregate demand of the \ac{dc} is modeled 
as: 
\begin{equation}\small
P^{\mathrm{DC}}(t) = P_b + P_{\rho}(t) + \sum_{j=1}^{N_w} p_j(t),
\label{eq:aggregate}
\end{equation}
where $P_b$ is the base load, comprising cooling and other quasi-constant infrastructure consumption, and $P_{\rho}(t)\sim\mathcal{N}(\mu_{\rho},\sigma_{\rho}^2)$ is the aggregate demand of the non-training workloads, i.e., inference, agentic, and CPU-intensive tasks, whose independent bursts average out into a nearly constant demand. $P_b$ and $\mu_{\rho}$ account for $30\%$ and $10\%$ of the facility consumption, respectively. 
Fig.~\ref{fig:pow_profile} illustrates the facility model through three profiles of equal mean demand that differ only in how the AI demand is distributed among workloads. The first, in which a single training job at $1.2$~Hz carries the entire demand, represents a frontier-scale facility or, equivalently, a facility fully coordinated by an adversary; it swings by 58\% of its mean peak-to-peak, including the \ac{psu} overshoot, and falls to about 80\% of it in every idle phase, in line with~\cite{jimenezruiz2026datacentre,hao_paper}. The second, in which all workloads are spread uniformly over $0.5$-$5$~Hz, reflects the conventional view of a multi-tenant facility: although composed of many square waves of different frequencies and duty ratios, it remains largely flat, with a swing of 12\%. The third, however, which is the most representative of an actual multi-tenant facility, tells a different story. With three workloads at $0.9$, $1.2$, and $1.4$~Hz carrying 40\% of the facility demand and the remainder spread over the same band, the facility still swings by 29\%, even though no workload is coordinated. Multi-tenant facilities are not inherently benign, as large oscillations may emerge from normal operation alone, without an adversary. In all cases, individual workloads retain the features of the measured traces of Fig.~\ref{fig:llm}, including the overshoot reproduced by the conversion stage, while the parameters used are listed in Table~\ref{tab:params}.

\begin{table}[t]
\caption{Workload and conversion-stage parameters. Bracketed entries are
uniform ranges; shares are of the facility consumption.}
\vspace{-2mm}
\label{tab:params}
\centering
\small
\setlength{\tabcolsep}{2pt}
\renewcommand{\arraystretch}{1.15}
\begin{tabular*}{\linewidth}{@{\extracolsep{\fill}}lclclc@{}}
\hline
\multicolumn{6}{c}{\rule{0pt}{2.6ex}\textit{Facility and Workload Model}} \\[0.3ex]
\hline
$P_{b}$ & $30\%$ & $\mu_{\rho}$, $\sigma_{\rho}$ & $10\%$, $0.1\mu_{\rho}$ & $\mu_{p}$, $\sigma_{p}$ & $0$, $0.5$ \\
$[\underline{f}_{0},\overline{f}_{0}]$ & $[0.5,5]$~Hz & $[\underline{r},\overline{r}]$ & $[0.2,0.8]$ & $\mu_{\Delta}$, $\sigma_{\Delta}$ & $0.47$, $0.01$ \\
$\sigma_{\eta,\mathrm{up}}$ & $[0.02,0.05]$ & $\sigma_{\eta,\mathrm{down}}$ & $[0.01,0.03]$ & & \\
\hline
\multicolumn{6}{c}{\rule{0pt}{2.6ex}\textit{Conversion Stage}} \\[0.3ex]
\hline
\end{tabular*}\\[-0.1ex]
\begin{tabular*}{\linewidth}{@{\extracolsep{\fill}}lclclc@{}}
$k_{p,p}$, $k_{i,p}$ & $314$, $2.47\!\times\!10^{4}$ & $k_{p,u}$, $k_{i,u}$ & $126$, $3.95\!\times\!10^{3}$ & $T_{\rm c}$ & $0.16$~ms \\
\hline
\end{tabular*}
\vspace{-2mm}
\end{table}
%
%
%

The persistence of this oscillation follows from how workloads dephase. Any two workloads $j$ and $k$ drift apart at the rate $\Delta f_{jk}=|f_{0,j}-f_{0,k}|$, passing from constructive to destructive alignment within $1/(2\Delta f_{jk})$~seconds, irrespective of their initial phases $\varphi_j$. For a fleet whose iteration rates are distributed uniformly over a band of width $B=\overline{f}_0-\underline{f}_0$, the phase coherence
\begin{equation}\footnotesize
    R(t) = \big|\textstyle\sum_{j} e^{\,\mathrm{j}2\pi f_{0,j}t}\big|/N_{w} \;\longrightarrow\; \left|\sin(\pi B t)/(\pi B t)\right|
    \label{eq:decay}
\end{equation}
first vanishes at $t=1/B$, i.e., after about $0.22$~s for the band considered here, and thereafter fluctuates around the finite-fleet floor $1/\sqrt{N_{w}}$. A dispersed fleet thus loses any initial alignment within a single iteration, which explains the flatness of the second profile. This mechanism, however, relies on the demand being spread over many workloads. When a few workloads carry a large share of it, as in the third profile, the floor $1/\sqrt{N_w}$ is no longer negligible, and the dominant workloads, separated by only $0.2$-$0.5$~Hz, drift back into alignment every $1/\Delta f_{jk}$~seconds, so that the oscillation recurs instead of decaying.

\begin{figure}[t]
    \centering
    \vspace{-2mm}
    \includegraphics[width=0.8\linewidth]{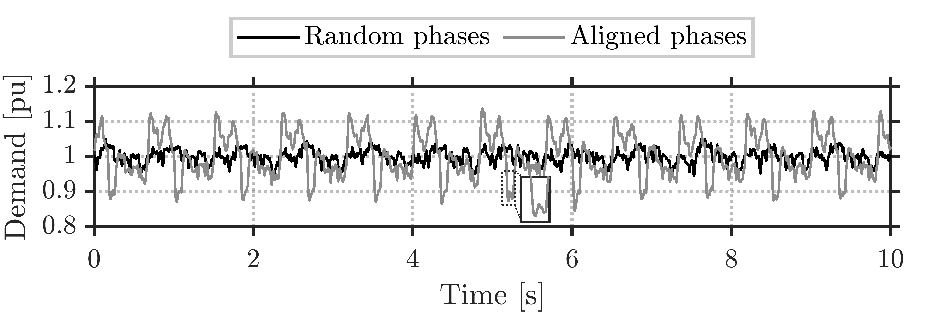}
    \vspace{-2mm}
\caption{Facility demand with harmonically related workloads under random and aligned initial phases.}
\label{fig:pow_phase}
\vspace{-2mm}
\end{figure}

Dephasing, moreover, cannot occur at all when workloads share identical or harmonically related iteration rates. Writing $f_{0,j}=m_{j}f_{r}$, with $m_{j}\in\mathbb{N}$ and $f_{r}$ a common base rate, the phase difference between any two workloads,
\begin{equation}\small
    \theta_{jk}(t) = 2\pi\left(m_{j}-m_{k}\right)f_{r}t + \left(\varphi_{j}-\varphi_{k}\right),
    \label{eq:frozen}
\end{equation}
returns to its initial value every $1/f_{r}$~seconds and reduces to the constant $\varphi_{j}-\varphi_{k}$ when $m_{j}=m_{k}$. The relative positions of such workloads therefore remain frozen in time, so that their initial phases alone determine how their demands combine, and hence the oscillatory energy and shape of the facility profile. This is illustrated in Fig.~\ref{fig:pow_phase}, where 50\% of the capacity is dispersed randomly across the $[\underline f_0,\overline f_0]$ band, 25\% is concentrated at $1.2$~Hz, and the remainder occupies harmonics of that frequency. The two traces share background workloads, duty ratios, and noise realizations, and differ only in the initial phases of the harmonically related workloads. When these phases are staggered, the oscillations of the workloads largely cancel and the facility remains comparatively flat. When they are aligned, however, the same facility, drawing the same average power from the same workloads, oscillates even more strongly than the uncoordinated multi-tenant profile of Fig.~\ref{fig:pow_profile}, solely because of when its workloads started. Phase alignment thus offers a powerful lever over the facility demand, one that an adversary can exploit to align the workloads and that the proposed mitigation exploits in turn to interleave them, under both adversarial and naturally arising synchronization.

\vspace{-2mm}
\section{Grid Vulnerability and Impact Assessment}
\label{sec:impact_assessment}
Since a sustained oscillation can be triggered by an adversary at will or arise naturally, two questions follow: which locations are most vulnerable to it, and how it propagates through the grid and interacts with regional assets such as generators. To address them, a modified IEEE 68-bus system hosting a \ac{dc} modeled as in Sect.~III is considered, with the impact bounded by the attacker capabilities of Sect.~II. All 16 generators are equipped with primary and secondary frequency control, capturing the regulating response of the grid itself. The assessment combines small- and large-signal analyses: the former reveals how the \ac{dc} interacts with the system modes depending on its location, whereas the latter captures how the generator controllers, driven far from equilibrium by MW-scale forcing, reshape this response.

\vspace{-3mm}
\subsection{Small-Signal Vulnerability Screening}
\vspace{-1mm}
To identify the most vulnerable buses of the network, the interaction of the \ac{dc} with the system modes, and its effect on the small-signal behavior of the grid, must be assessed as a function of its location. To this end, a \ac{dc} with an average demand of 420~MW is considered, and the small-signal model of the benchmark is derived with the facility demand $P_k$ at its site~$k$ as input and, as outputs, the states $\chi_m$ observed at locations~$m$, such as the speed of a generator or the voltage of a bus. The \ac{dc} is then sited at each bus in turn, and for every oscillatory mode $\lambda_i=-\sigma_i+\mathrm{j}\omega_i$ and every pair $(k,m)$, the following residue is evaluated:
\begin{equation}\small
  R_i(k,m) = \left(c_m\,v_i\right)\left(w_i^{\top} b_k\right)
  \label{eq:residue}
\end{equation}
where $v_i$ and $w_i$ are the right and left eigenvectors of mode~$i$, with $w_i^{\top}v_i=1$, and $b_k$ and $c_m$ are the input vector and output row associated with $P_k$ and $\chi_m$, respectively. The residue thus combines how strongly site~$k$ excites mode~$i$, through $w_i^{\top} b_k$, with how visible the mode is at location~$m$, through $c_m v_i$.
\begin{figure}[t]
  \centering
  \includegraphics[width=0.8\columnwidth]{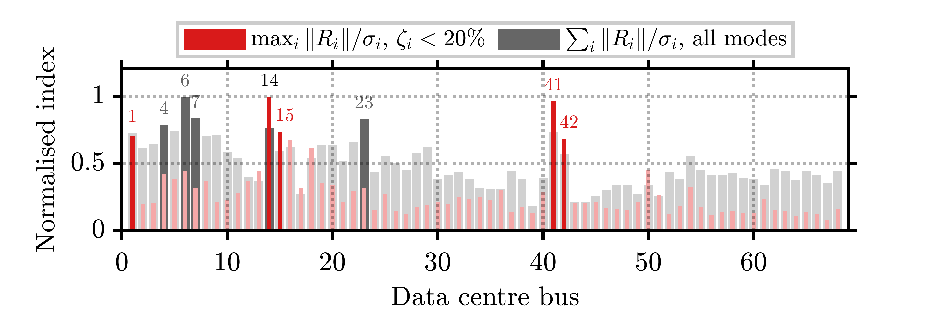}
  \vspace{-2mm}
  \caption{Residue-based screening of the buses at which a \ac{dc} interacts most strongly with the system modes.}
  \label{fig:energy}
  \vspace{-1mm}
\end{figure}


The results are depicted in Fig.~\ref{fig:energy}, obtained for $m\in\mathcal{G}$, with $\mathcal{G}$ the set of generators and $\chi_m$ their speeds, and with the inertia of G7 reduced to 70\% of its original value, following the paradigm of~\cite{hao_paper}. The buses are ranked under two criteria. The first, adopted in~\cite{hao_paper}, deems a bus vulnerable when the \ac{dc} interacts strongly with a single lightly damped mode, i.e., $k^\star=\argmax_k \max_{i:\,\zeta_i<\zeta_{\max}} \|\boldsymbol{R}_i(k)\|/|\sigma_i|$, where $\boldsymbol{R}_i(k)=[R_i(k,m)]_{m\in\mathcal{G}}$, $|\sigma_i|$ and $\zeta_i$ are the decay rate and damping ratio of mode~$i$, and $\zeta_{\max}=20\%$. Under this criterion, bus~14 ranks highest, although it hosts one of the highest-inertia generators of the system. The oscillation imposed by the \ac{dc}, however, excites several modes at once, and its effect is better captured by aggregating all oscillatory modes, i.e., $k^\star=\argmax_k \sum_i \|\boldsymbol{R}_i(k)\|/|\sigma_i|$. Under this criterion, the ranking changes substantially: bus~14 is the only bus that remains among the most vulnerable, while generator buses~4, 6, and~7 become the most prominent. Buses~4 and~7 are electrically adjacent and host lower-inertia machines, notably G7, whose inertia was reduced to further illustrate the potency of the attack, as bus~7 already ranks highest under this criterion in the unmodified benchmark. Bus~23 is similarly vulnerable owing to its electrical proximity to bus~7.

The aggregate criterion does not reveal the forcing frequencies at which this interaction peaks, nor the locations these frequencies excite most. To capture both, the small-signal model is evaluated with a single \ac{dc} connected at each bus, yielding for every site~$k$ and location~$m$ the transfer function:
\begin{equation}\small
H_{k,m}(s) = \Delta \chi_m(s)/\Delta P_{k}(s) = \textstyle\sum_i R_{i}(k,m)/(s-\lambda_i), \label{eq:tf}
\end{equation}
where $\Delta P_k$ is the demand of the \ac{dc} connected at bus~$k$, $\Delta\chi_m$ is a state of interest at location~$m$, e.g., the speed deviation of a generator or the voltage of a bus, and $R_{i}(k,m)$ is the residue of mode~$i$ for this input-output pair, obtained through partial fraction decomposition. In this section, $\chi_m$ denotes the speed deviation of generator~$m$.

\begin{figure}[!t]
  \centering \includegraphics[width=\columnwidth]{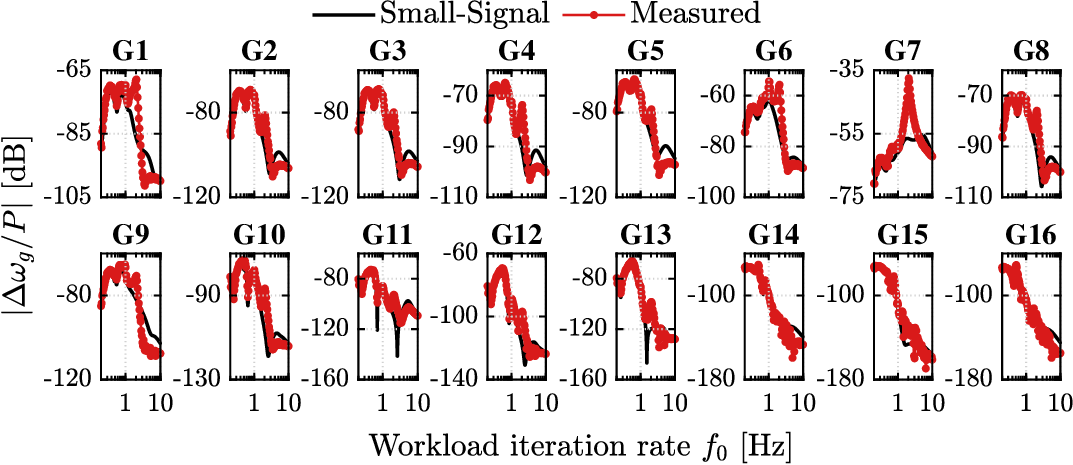}
  \vspace{-4mm}
  \caption{Frequency response $H_{23,m}(\mathrm{j}\omega)$ of each generator: small-signal model (black) and time-domain measurement (red).}
  \label{fig:bode}
  \vspace{-1mm}
\end{figure}

\begin{figure}[t]
  \centering
  \includegraphics[width=0.8\columnwidth]{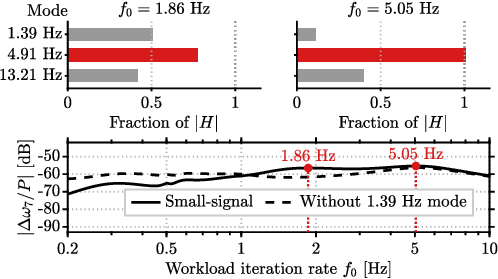}
  \vspace{-1mm}
  \caption{Modal contributions to $|H_{23,7}(\mathrm{j}\omega)|$ at its two resonance peaks.}
  \label{fig:residue}
  \vspace{-1mm}
\end{figure}

As an example, placing the \ac{dc} at bus~23, one of the most vulnerable sites identified above, and evaluating \eqref{eq:tf} for every generator~$m$ yields the frequency responses illustrated in Fig.~\ref{fig:bode}, which reveal the generators most susceptible to resonance and the workload frequencies that excite them. G7 exhibits the highest gain, with two resonance peaks near 1.86~Hz and 5.05~Hz, as expected from its lower inertia and its electrical proximity to the \ac{dc}. Notably, the peak at 1.86~Hz coincides with no mode of the network, but results from the interaction of several modes. To demonstrate this, the modal term $R_i(23,7)/(\mathrm{j}\omega-\lambda_i)$ of each mode is evaluated at the two peak frequencies, and its magnitude relative to $|H_{23,7}(\mathrm{j}\omega)|$ indicates its contribution to the response. As shown in Fig.~\ref{fig:residue}, the response at 1.86~Hz is shaped jointly by the 1.39~Hz, 4.91~Hz, and 13.21~Hz modes, with the removal of the 1.39~Hz mode suppressing the peak altogether, whereas at 5~Hz it is dominated by the 4.91~Hz mode.

To identify the \ac{dc} site, affected location, and waveform that jointly maximize the excitation, the location- and frequency-dependent sensitivity of \eqref{eq:tf} is weighted by the workload controllability measured in Sect.~II. This yields the worst-case location, both as an adversarial target and for siting a \ac{dc}, since any oscillation, benign or malicious, would manifest there most severely:
\begin{subequations}
\label{eq:opt}
{\small
\begin{align}
  \max_{k,\,m,\,f_0,\,r} \quad & \big|H_{k,m}(\mathrm{j}2\pi f_0)\big|\,\mathcal{P}(f_0,r) \label{eq:opt_obj}\\
  \text{s.t.} \quad & \mathcal{P}(f_0,r) = 2\,\gamma(f_0,r)\,\mu_{\Delta}\,\hat p\,r\,\lvert\operatorname{sinc}(r)\rvert, \label{eq:opt_P}\\
  & k \in \mathcal{B}, \quad m \in \mathcal{M}, \label{eq:opt_set}\\
  & 0.2 \le f_0 \le 5~\text{Hz}, \quad 0.2 \le r \le 0.8, \label{eq:opt_fr}
\end{align}}
\end{subequations}
where $\mathcal{B}$ and $\mathcal{M}$ are the sets of candidate \ac{dc} sites and observed locations, respectively. $\mathcal{P}(f_0,r)$ is the fundamental amplitude of the aligned workload fleet, with aggregate up-phase power $\hat p$, iteration rate $f_0$, duty ratio $r$, and $\operatorname{sinc}(x)=\sin(\pi x)/(\pi x)$. To account for limited controllability of the workload, this ideal amplitude is weighted by the fidelity index $\gamma(f_0,r)\in[0,1]$, the correlation between the measured and commanded GPU power at rate $f_0$ and duty ratio $r$, with $\epsilon_{\mathrm{rms}}=\sqrt{2(1-\gamma)}$ the error shown in Fig.~\ref{fig:heat}. Solving this problem for $\mathcal{M}=\mathcal{G}$, the set of generators, with \acp{dc} excluded from generator buses, yields $(k^\star,m^\star,f_0^\star,r^\star)=(42,\,\mathrm{G}15,\,0.78~\text{Hz},\,0.53)$, which departs from the bus-23 example once the excitation is confined to what the attacker can faithfully impose.
%
%
%
\vspace{-4mm}
\subsection{Large-Signal Response to Forced Oscillations}
\label{sec:large_signal}
\vspace{-1mm}
The small-signal model is valid only for small perturbations around the equilibrium, a condition violated here owing to the scale of the \ac{dc} oscillation, which may therefore alter the previous findings. To examine this, the large-signal frequency response of the test system is measured through nonlinear time-domain simulations, with the \ac{dc} placed at bus~23 and its consumption given by the single-workload profile of Fig.~\ref{fig:pow_profile}. At each iteration rate $f_0$, the measured response is $\hat{H}_{k,m}(\mathrm{j}\omega_0)=\Delta\hat{\chi}_m(\omega_0)/\hat{P}_k(\omega_0)$, where $\Delta\hat{\chi}_m(\omega_0)$ and $\hat{P}_k(\omega_0)$ are the complex fundamental components of the observed state and of the \ac{dc} consumption at $\omega_0=2\pi f_0$, extracted over the last $M$ periods of the steady-state response.

\begin{figure}[t]
  \centering
  \includegraphics[width=0.8\columnwidth]{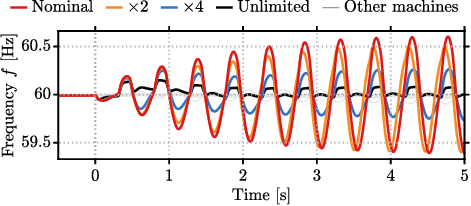}
  \vspace{-2mm}
  \caption{G7 frequency under scaled AVR and PSS limits.}
  \label{fig:limiter}
  \vspace{-1mm}
\end{figure}

The results, shown in Fig.~\ref{fig:bode}, deviate markedly from the small-signal prediction. The response of most generators is considerably larger than in the linearized case, and the most critical excitation shifts toward a new resonance at G7, which peaks at 2.06~Hz, about 19~dB above the small-signal response. This resonance lies beyond the reach of the linear model, as it originates in the nonlinear elements of the generator controls: the oscillation drives the limiters of the power system stabilizers (PSSs) and automatic voltage regulators (AVRs) into saturation, which, for G7, occurs at demand oscillations of only about 5 and 70~MW peak-to-peak for the PSS and AVR, respectively, in the unmodified benchmark, and 2 and 10~MW under the reduced inertia adopted here to expose this mechanism, far below the hundreds of MW a single \ac{dc} can swing. Once saturated, these controllers lose authority precisely when it is most needed, and the oscillation they are meant to damp is amplified instead~\cite{observed_8930066}. Fig.~\ref{fig:limiter} isolates this mechanism: as the AVR and PSS limits of all generators are progressively relaxed, the oscillation of G7 diminishes, and once the limits are removed, it falls back to the level predicted by the small-signal model. Anti-windup offers little help, as the PSS contains no integrator and active limiters clip the control action regardless; indeed, adding it to the AVR altered the response by less than 1\%. This large-signal effect can be integrated into the vulnerability assessment by replacing $H_{k,m}$ in \eqref{eq:opt_obj} with the measured response $\hat{H}_{k,m}$, valid in the vicinity of the measured excitation amplitude.
The resulting oscillation also differs in nature from those usually encountered in power systems. Sustained oscillations typically signal an intrinsic deficiency of the network, such as insufficient damping, under which the equilibrium loses stability and the trajectory settles onto a limit cycle, an isolated periodic orbit of the autonomous dynamics. Here, in contrast, the equilibrium remains asymptotically stable, and the oscillation is imposed by the periodic workload, which renders the system non-autonomous and steers its trajectory onto a persistent, bounded periodic orbit. The \ac{dc} thus induces a forced periodic orbit, one that exhibits the observable features of a limit cycle without being generated by the dynamics of the network. An adversary can impose it at will, simply by tuning the iteration rate to the large-signal resonance of the network. The consequences are the same as those of a conventional limit cycle, namely sustained oscillations that may lead to generator curtailment or tripping. 


\vspace{-4mm}
\subsection{Stability Analysis of Periodic Solutions}
\vspace{-1mm}
Floquet theory allows the impact of the oscillation on the small-signal dynamics of the system, and in particular on modal damping, to be evaluated. The power system is described by the differential-algebraic equations:
\begin{equation}\small
    \bfg x^{\prime}=\bfg f(\bfg x,\bfg y,\bfg u), \qquad \bfg 0=\bfg g(\bfg x,\bfg y,\bfg u),
    \label{eq:f}
\end{equation}
where $\bfg x\in\mathbb{R}^{n}$ is the state vector, $\bfg y\in\mathbb{R}^{n_y}$ the vector of algebraic variables, and $\bfg u\in\mathbb{R}^{n_u}$ the input vector, including \ac{dc} consumption. Linearizing around a stationary equilibrium $(\bfg x_0,\bfg y_0,\bfg u_0)$, where $\bfg f=\bfg 0$ and $\bfg g=\bfg 0$, and eliminating the algebraic variables, assuming $\bfg g_{\bfg y}$ nonsingular, yields:
\begin{equation}\small
    \Delta\bfg x^{\prime}=\bfg A\,\Delta\bfg x+\bfg B\,\Delta\bfg u,
\end{equation}
with $\bfg A=\bfg f_{\bfg x}-\bfg f_{\bfg y}\bfg g_{\bfg y}^{-1}\bfg g_{\bfg x}$ and $\bfg B=\bfg f_{\bfg u}-\bfg f_{\bfg y}\bfg g_{\bfg y}^{-1}\bfg g_{\bfg u}$ evaluated at the equilibrium. Under a periodic workload, however, the system no longer settles at an equilibrium but on a periodic orbit $(\bar{\bfg x},\bar{\bfg y},\bar{\bfg u})(t)=(\bar{\bfg x},\bar{\bfg y},\bar{\bfg u})(t+T)$, where $T$ is the period of the forcing. Linearizing along this solution, with $\bfg g_{\bfg y}$ nonsingular along it, yields the \ac{ltp} system $\Delta\bfg x^{\prime}=\bfg A(t)\,\Delta\bfg x+\bfg B(t)\,\Delta\bfg u$, with $\bfg A(t)=\bfg A(t+T)$ and $\bfg B(t)=\bfg B(t+T)$, whose periodicity originates from the workload in $\bar{\bfg u}(t)$~\cite{leine2004dynamics}. Since stability concerns perturbations of the state under the same forcing, i.e., $\Delta\bfg u=\bfg 0$, it is governed by $\bfg A(t)$ alone, and the free response is $\Delta\bfg x(t)=\bfg\Phi(t,t_0)\Delta\bfg x(t_0)$, where the state transition matrix $\bfg\Phi\in\mathbb{R}^{n\times n}$ satisfies:
\begin{equation}\small
    \bfg\Phi^{\prime}(t,t_0)=\bfg A(t)\bfg\Phi(t,t_0), \qquad \bfg\Phi(t_0,t_0)=\bfg I.
\end{equation}
Owing to the periodicity of $\bfg A(t)$, $\bfg\Phi(t+T,t_0)=\bfg\Phi(t,t_0)\bfg\Phi(t_0+T,t_0)$, so that the evolution over any number of periods is governed by the \textit{monodromy} matrix $\bfg\Phi(t_0+T,t_0)$, the basis of Floquet theory~\cite{leine2004dynamics}:
\begin{theorem}
    The periodic solution is asymptotically stable if all eigenvalues of the monodromy matrix lie in the open unit disk, $\lambda\{\bfg\Phi(t_0+T,t_0)\}\subset\mathcal{D}_0, \quad \mathcal{D}_0=\{z\in\mathbb{C} : |z|<1\}.$
\end{theorem}
These eigenvalues, the Floquet multipliers $\mu_i$, are the discrete-time counterparts of the Floquet exponents $\nu_i=\tfrac{1}{T}\ln\mu_i=-\sigma_i+\mathrm{j}\omega_i$, where $\sigma_i$ is uniquely defined and directly comparable with the decay rate of the stationary eigenvalues, whereas $\omega_i$ is determined only up to integer multiples of $2\pi/T$. The generator limiters render \eqref{eq:f} non-smooth, but since they saturate signals between control blocks rather than states, the vector field remains continuous at their switching instants, and only the Jacobian, and hence $\bfg A(t)$, changes abruptly, so that the saltation matrix reduces to the identity~\cite{hybrid_saltatation}.

\begin{figure}[t]
  \centering
  \begin{subfigure}[b]{0.38\columnwidth}
    \centering
    \includegraphics[width=\linewidth]{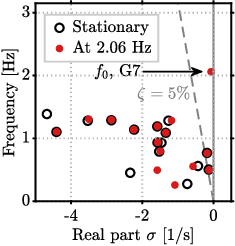}
    \caption{}
    \label{fig:floquet_plane}
  \end{subfigure}\hfill
  \begin{subfigure}[b]{0.38\columnwidth}
    \centering
    \includegraphics[width=\linewidth]{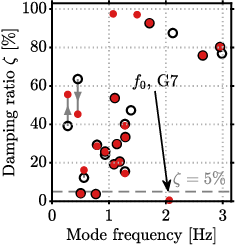}
    \caption{}
    \label{fig:floquet_damping}
  \end{subfigure}
  \vspace{-2mm}
  \caption{Stationary eigenvalues and Floquet exponents of the periodic orbit at 2.06~Hz: (a) complex plane, (b) damping ratio of each mode.}
  \label{fig:floquet}
  \vspace{-1mm}
\end{figure}

To assess the impact of this oscillation, Floquet analysis is applied to the same system driven by the single workload at 2.06~Hz, with the deterministic profile used so that the orbit is exactly periodic. The resulting exponents are compared in Fig.~\ref{fig:floquet} with the eigenvalues of the stationary equilibrium examined in Sect. IV-A. Although the periodic orbit remains asymptotically stable, several modes become less damped along it, e.g., the damping of the 0.45~Hz mode decreases from 64\% to 45\%. More importantly, a practically undamped mode emerges at the forcing frequency, with a damping ratio of about 0.5\%, which is continuously excited by the workload, corroborating the large-signal resonance, whereas no comparably damped mode exists in this band at the stationary equilibrium. In order to verify these results, a harmonic analysis of each mode along the orbit is performed, confirming that the energy of the matched modes is concentrated at the depicted frequencies (99-100\%), and that of the emerging mode at 2.06~Hz.
\vspace{-3mm}
\section{Workload-Level Detection and Mitigation Using Signal-Energy Decomposition}
\vspace{-1mm}
Building on the modeling and analyses of Sect.~III and~IV, this section develops detection and mitigation mechanisms that limit the grid-level consequences of adverse \ac{dc} demand dynamics. Both rely on signal energy of the workload power profiles: detection uses it to attribute irregularities in the facility demand to the responsible workloads, while mitigation uses it to coordinate their execution to strengthen the resilience of the \ac{dc} against malicious or benign synchronization. 
\vspace{-3mm}
\subsection{Detection of Irregular Workload Patterns}
\begin{figure}[t]
\centering
\includegraphics[width=0.8\linewidth]{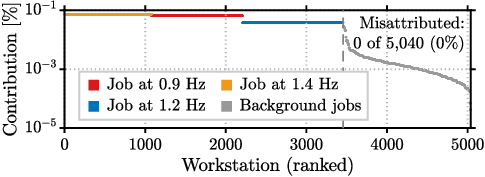}
\vspace{-2mm}
\caption{Ranked contribution of the 5040 nodes to the facility oscillation.}
\vspace{-1mm}
\label{fig:detect}
\end{figure}

To trace irregularities in the facility demand, such as oscillations or impulses, back to the responsible workstations, the contribution of each workload is quantified through the signal energy of its power trace. With a slight abuse of notation, $p_i$ henceforth denotes the power trace of any workload~$i\in\mathcal{W}$ or of any of its compute nodes, where $\mathcal{W}$ is the set of all workloads hosted by the facility, including the non-training ones. For each workload~$i$, the facility demand \eqref{eq:aggregate} is thus decomposed as:
\begin{equation}\small
    \tilde{P}^{\mathrm{DC}}(t)=\tilde{p}_i(t)+\tilde{q}_i(t), \qquad q_i(t)=\sum_{j\in\mathcal{W},\,j\neq i}p_j(t),
\end{equation}
where $q_i$ is the consumption of the remaining fleet and $\tilde{(\cdot)}$ denotes a signal with its mean over an observation window $T_w$ removed, so that the base load $P_b$ does not enter. Defining the energy of the deviations over $T_w$ of the facility demand and of the remaining fleet,
\begin{equation}\small
    E_{\mathrm{DC}}=\int_{T_w}\big(\tilde{P}^{\mathrm{DC}}(t)\big)^2dt, \qquad E_{q_i}=\int_{T_w}\tilde{q}_i^{\,2}(t)\,dt,
\end{equation}
the marginal energy contribution of workload~$i$ to the facility demand, $\Delta E_i=E_{\mathrm{DC}}-E_{q_i}$, becomes:
\begin{equation}\small
    \Delta E_i=\underbrace{\int_{T_w}\tilde{p}_i^{\,2}\,dt}_{\text{individual energy}}+\underbrace{2\int_{T_w}\tilde{p}_i\,\tilde{q}_i\,dt}_{\text{interaction with the fleet}}.
\end{equation}
The marginal energy contribution thus quantifies the responsibility of each workload for the deviations of the facility demand, with the sign of the interaction term indicating whether the deviations of the workload reinforce or attenuate those of the remaining fleet. For oscillatory phenomena, however, the danger arises mainly within specific frequency bands, as shown in Sect. IV, which this expression cannot resolve. To this end, let $\tilde{\mathcal{P}}_i(f)$ and $\tilde{\mathcal{Q}}_i(f)$ denote the Fourier transforms of $\tilde{p}_i$ and $\tilde{q}_i$ over $T_w$. By Parseval's theorem, $\int_{T_w}\tilde{p}_i^{\,2}\,dt=\int|\tilde{\mathcal{P}}_i(f)|^2df$ and $\int_{T_w}\tilde{p}_i\tilde{q}_i\,dt=\int\Re\{\tilde{\mathcal{P}}_i(f)\tilde{\mathcal{Q}}_i^{*}(f)\}\,df$, so that the frequency-resolved contribution: 
{\small
\begin{align}
    \Delta E_i(f)&=|\tilde{\mathcal{P}}_i(f)|^2+2\Re\{\tilde{\mathcal{P}}_i(f)\,\tilde{\mathcal{Q}}_i^{*}(f)\}\nonumber\\
    &=|\tilde{\mathcal{P}}_i|^2+2|\tilde{\mathcal{P}}_i||\tilde{\mathcal{Q}}_i|\cos\!\big(\angle\tilde{\mathcal{P}}_i-\angle\tilde{\mathcal{Q}}_i\big)
\label{eq:delta_e_i}
\end{align}}satisfies $\Delta E_i=\int\Delta E_i(f)\,df$. Restricting the integral to a critical band $\mathcal{F}$, e.g., around the resonance of the network, yields the band-limited contribution $\Delta E_i(\mathcal{F})=\int_{\mathcal{F}}\Delta E_i(f)\,df$, which isolates the most harmful part of the facility spectrum. The proposed detection mechanism thus operates as follows: the workload traces and the facility demand are monitored over a sliding window $T_w$, and whenever an irregularity emerges, $\Delta E_i$ or $\Delta E_i(\mathcal{F})$ is evaluated for every workload, flagging those with the largest positive contributions as responsible, while those with negative ones counteract the irregularity, whether impulsive, oscillatory or of any other form. This is illustrated in Fig.~\ref{fig:detect} for the third profile of Fig.~\ref{fig:pow_profile}, whose jobs are split into 5040 nodes of 50~kW, i.e., the training demand of a facility of about 420~MW, that run in lockstep: ranked by their contribution $\Delta E_i$ over the bands detected in the facility spectrum, all 3455 nodes of the three large jobs precede every background node, so that none is misattributed. The scheme thus attributes the irregularity to workloads rather than inferring intent, which is left to the policies of the operator.   

This detection mechanism lays the foundation for dedicated policies through which \acp{dc} can safeguard both the facility and the grid. Such policies may specify the actions taken against workloads that cause substantial deviations in the facility demand, such as restricting their resources, imposing monetary penalties, or curtailing them altogether. Regulating training workloads, however, is inherently challenging, since the same execution patterns may arise naturally, without the presence of an adversary, and the two are difficult to distinguish. As a first line of defense, operators should therefore require users to declare their intention to train a priori. Compliance can then be verified, since undeclared training is exposed by correlating workstation consumption with the workload model of Sect.~III. Under these provisions, training is confined to declared workloads, which, even if they conceal a stealthy adversary, fall under the full control of the operator, who can either restrict them or, as shown next, mitigate their impact.

\vspace{-3mm}
\subsection{Mitigation via Frequency-Aligned Workload Interleaving}
\vspace{-1mm}
To mitigate harmful execution patterns, whether arising naturally or from adversarial scheduling, the proposed mechanism exploits the condition identified in Sect.~III under which natural dephasing fails: once workloads share a common base frequency or its harmonics, their relative positions remain frozen in time, and their phase differences alone determine the shape of the facility demand. The mitigation proceeds in two stages. 
In the first stage, the workloads are brought onto a common base frequency $f_b$, dictated by the slowest among them, or onto its harmonics, by prolonging their idle phases, since a workload cannot be accelerated. The slowed profile $p_i'$, where $(\cdot)'$ denotes a quantity after the frequency change, is built directly from the measurement snapshot of length $T_w$: each compute phase is kept as recorded, while the idle phase that follows it is extended by $\Delta T_i=1/f_i'-1/f_i$ by repeating its own recorded samples, where $f_i$ and $f_i'$ are the original and new frequencies of workload~$i$. The new frequencies and the placement of the workloads are thus determined in a single step, without waiting for the hardware response. Both $f_b$ and $f_i'$ may themselves be optimized, balancing the state of the grid against the clients' quality of service, as lower rates prolong their execution.

In the second stage the workloads are placed in time by one-time delays $\bfg\tau=[\tau_i]_{i\in\mathcal{C}}$, applied by extending a single low-power phase of each controlled workload $i\in\mathcal{C}$. By the time-shift property, a delayed workload has the spectrum $\tilde{\mathcal{P}}_i'(f)e^{-\mathrm{j}2\pi f\tau_i}$, where $\tilde{\mathcal{P}}_i'$ is the spectrum of $\tilde{p}_i'$. On the common base frequency the controlled workloads occupy only the harmonics $hf_b$, $h\in\mathbb{N}$, so by Parseval's theorem, and up to a constant, minimizing the oscillatory energy of the aggregate amounts to:
{\small
\begin{align}
    \min_{\bfg\tau}\;\; & \sum_{h\geq1} w_h\Big|\tilde{\mathcal{U}}(hf_b)+\sum_{i\in\mathcal{C}}\tilde{\mathcal{P}}_i'(hf_b)\,e^{-\mathrm{j}2\pi hf_b\tau_i}\Big|^2 \label{eq:mitig}\\
    \text{s.t.}\;\; & 0\leq\tau_i<1/f_i', \qquad i\in\mathcal{C}, \nonumber
\end{align}}
where $\tilde{\mathcal{U}}$ is the spectrum of the uncontrolled workloads and $w_h=1+W_c\,\mathbf{1}_{\mathcal{F}}(hf_b)$ weights the critical band $\mathcal{F}$. 

To solve this problem, demand is split, as in the detection metric, into workload~$i$ and remaining fleet, 
{\small$\tilde{\mathcal{Q}}_i'(hf_b)=\tilde{\mathcal{U}}(hf_b)+\sum_{\ell\in\mathcal{C},\,\ell\neq i}\tilde{\mathcal{P}}_\ell'(hf_b)\,e^{-\mathrm{j}2\pi hf_b\tau_\ell}$}, which yields: {\small $ J(\bfg\tau) ={} \sum_{h\geq1} w_h\Big(|\tilde{\mathcal{Q}}_i'(hf_b)|^2+|\tilde{\mathcal{P}}_i'(hf_b)|^2\Big) \nonumber
    + 2\,\Re\Big\{\sum_{h\geq1} w_h\,\tilde{\mathcal{P}}_i'(hf_b)\,\tilde{\mathcal{Q}}_i'^{*}(hf_b)\,e^{-\mathrm{j}2\pi hf_b\tau_i}\Big\}.$}

Only the last term depends on $\tau_i$, and since it is the cross-correlation of workload~$i$ with the remaining fleet, a single inverse FFT of $w_h\tilde{\mathcal{Q}}_i'\tilde{\mathcal{P}}_i'^{*}$ can evaluate it for every candidate delay, its $n$-th sample corresponding to a delay of $n$ samples; the optimal delay of workload~$i$, given the others, is the one minimizing its real part over one period of the workload. This exact and inexpensive step motivates solving~\eqref{eq:mitig} by coordinate descent. In each pass, every workload in turn is moved to its optimal delay with the others fixed. The passes are repeated until the relative decrease of $J$ falls below a tolerance~$\varepsilon$. To avoid a poor local minimum, the descent is run $R$ times, first in decreasing order of power and then in random orders, and the delays with the lowest $J$ are kept, as summarized in Alg.~\ref{alg:mitig}. Notably, the methodology also extends to the synchronization of workloads across geographically distributed facilities, a case examined in Sect.~VI.
\begin{algorithm}[t]
\caption{Workload interleaving by coordinate descent}
\small{
\label{alg:mitig}
\begin{algorithmic}[1]
\Require $p_i(t)$, $i\in\mathcal{C}$, and $\tilde{\mathcal{U}}$ over $T_w$; $f_b$; $w_h$; $\varepsilon$; $R$
\Ensure $f_i'$, $\tau_i$, $i\in\mathcal{C}$
\For{$i\in\mathcal{C}$}
    \State $p_i' \gets p_i$ slowed to the harmonic $f_i'$ of $f_b$ by extending its low phase by $\Delta T_i$
    \State $\tilde{\mathcal{P}}_i' \gets \mathrm{FFT}\{p_i'\}$
\EndFor
\For{$r=1,\dots,R$}
    \State order $\gets$ decreasing power if $r=1$, else random
    \State $\tilde{\mathcal{P}}^{\mathrm{AI}\prime} \gets \tilde{\mathcal{U}}$;\; $\tilde{\mathcal{D}}_i' \gets 0$, $i\in\mathcal{C}$
    \Repeat
        \For{$i$ in order}
            \State $\tilde{\mathcal{Q}}_i' \gets \tilde{\mathcal{P}}^{\mathrm{AI}\prime}-\tilde{\mathcal{D}}_i'$
            \State $\tau_i \gets \argmin_{\tau_i\in[0,1/f_i')} J(\bfg\tau)$
            \State $\tilde{\mathcal{D}}_i' \gets \tilde{\mathcal{P}}_i' e^{-\mathrm{j}2\pi f\tau_i}$;\; $\tilde{\mathcal{P}}^{\mathrm{AI}\prime} \gets \tilde{\mathcal{Q}}_i'+\tilde{\mathcal{D}}_i'$
        \EndFor
    \Until{$J_{\kappa-1}-J_\kappa<\varepsilon J_{\kappa-1}$},  with $\kappa$ the pass index
    \State keep $\bfg\tau$ if $J$ is the lowest so far
\EndFor
\State \Return $f_i'$, $\tau_i$, $i\in\mathcal{C}$
\end{algorithmic}}
\vspace{-1mm}
\end{algorithm}

\begin{figure}[t]
  \centering
  \vspace{-3mm}
  \includegraphics[width=0.8\columnwidth]{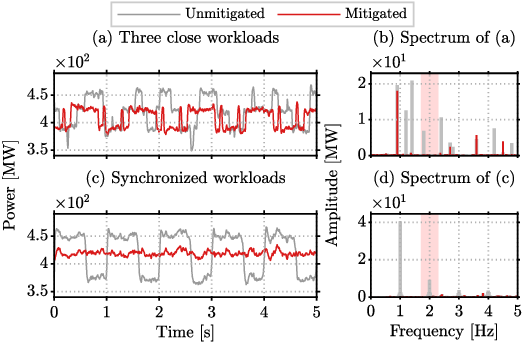}
  \vspace{-2mm}
  \caption{Facility demand before/after mitigation: (a) three workloads at 0.9, 1.2, and 1.4~Hz; (c) 100 in-phase workloads at 1~Hz; (b), (d) corresponding spectra, with the critical band (1.7-2.3~Hz) shaded.}
  \label{fig:mitig}
  \vspace{-1mm}
\end{figure}

\begin{figure}[t]
    \centering
    \vspace{-4mm}
    \includegraphics[width=0.8\columnwidth]{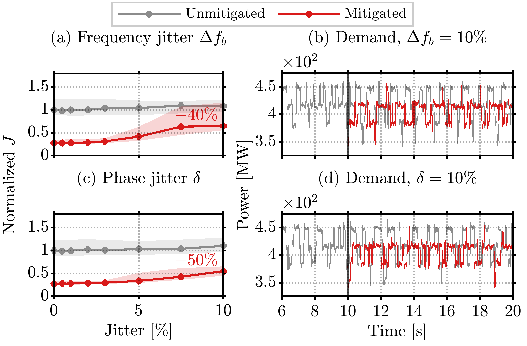}
    \vspace{-2mm}
\caption{Mitigation robustness to workload irregularities: frequency (a,b) and phase (c,d) jitter.}
\label{fig:jitter}
\vspace{-1mm}
\end{figure}

The performance of the mitigation is illustrated in Fig.~\ref{fig:mitig} for two cases. In the first, the three dominant workloads of the third profile in Fig.~\ref{fig:pow_profile} are slowed to 0.9~Hz and optimally delayed, reducing the RMS oscillation of the facility demand from 26.6 to 17.3~MW and its amplitude in the critical band by 88\%. The second considers the extreme case of adversarial synchronization, in which 100 workloads operate in phase at 1~Hz; optimally delayed, they cancel almost entirely, reducing the RMS oscillation from 30.7 to 4.6~MW. With $R=10$ runs and $\varepsilon=10^{-2}$, the coordinate descent terminates after 0.09 and 1.2~s, respectively, within 1.1\% of the optimum and 0.6\% of the best known solution, rendering the mechanism suitable for online operation. Workloads may, however, be irregular, with a per-iteration frequency that varies either directly (frequency jitter $\Delta f_b$) or through independent variations of their compute and idle phases (phase jitter $\delta$), causing them to drift out of phase even after being brought to a common base frequency. In this case, the delays are simply re-optimized periodically, here every eight periods, and, as shown in Fig.~\ref{fig:jitter}, the mitigation retains its full effect for jitter up to 3\% and still reduces the oscillation energy by 40\% to 50\% at 10\%.


\vspace{-3mm}
\section{Case Studies}
\begin{figure}[t]
    \centering
    \includegraphics[width=0.8\columnwidth]{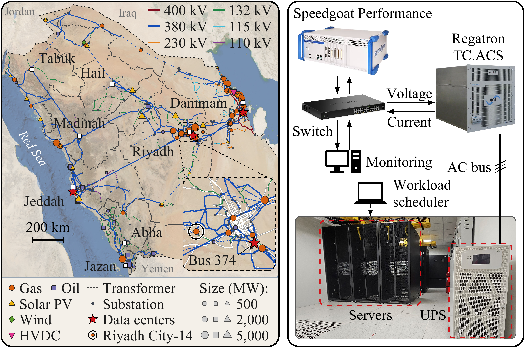}
  \caption{Surrogate Saudi transmission model \ac{phil} experiment.}
  \label{fig:saudi}
  \vspace{-5mm}
\end{figure}

The preceding analyses are validated in this section on a surrogate model of the Saudi Arabian transmission network with 799 buses and 73 synchronous generators, built from publicly available information~\cite{osm_saudi,sec2020annual} and shown in Fig.~\ref{fig:saudi}, simulated on a Speedgoat Performance target with a $10^{-4}$~s time step. In the first experiment, a \ac{phil} setup reproduces an attack by an adversary following the threat model of Sect.~II. The \ac{dc} is emulated by the laboratory computing infrastructure of Fig.~\ref{fig:saudi}, comprising 20 compute nodes, each with two AMD EPYC~9124 processors and a peak power of about 450~W, i.e., about 9~kW in total. As no sufficiently large GPU cluster was available, the power profiles of \ac{llm} training, generated by the model of Sect.~III, are reproduced on these CPU nodes by controlling their utilization. All nodes are time-synchronized via NTP and launched together, so that their power transitions align within 100~ms. The \ac{dc} is interfaced with the simulation through a Regatron amplifier, which supplies it with the simulated nodal voltage while its drawn current is measured and injected back into the network. In the second, a purely real-time simulation of the same network, with the facilities represented by the model of Sect.~III, demonstrates how the proposed mechanism suppresses the impact of synchronized \ac{llm} training across geographically distributed facilities.
%
%
\begin{figure}[t]
    \centering
    \includegraphics[width=0.8\columnwidth]{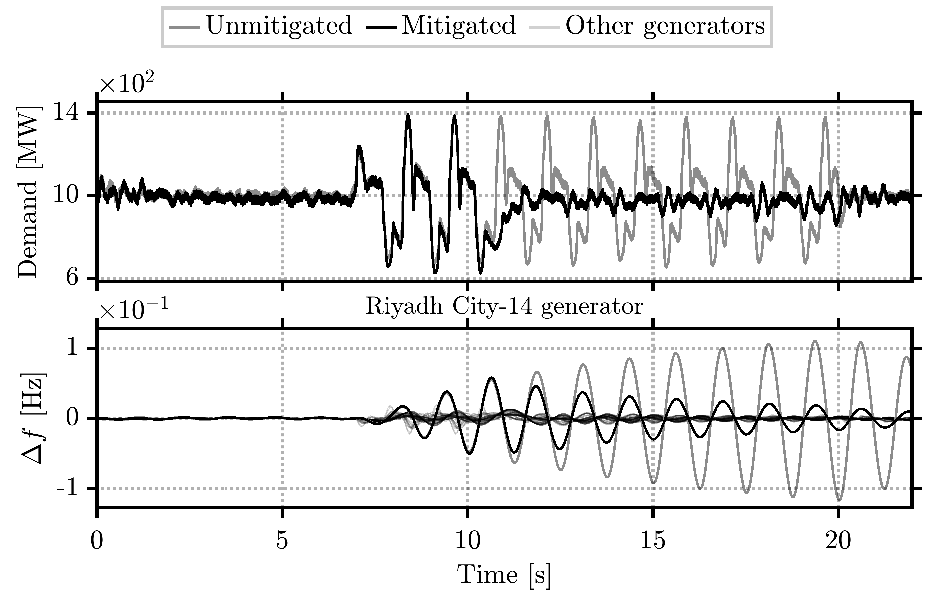}
    \vspace{-2mm}
    \caption{System response during the synchronization attack: \ac{dc} demand (top) and frequency of the Riyadh City-14 generator (bottom).}
    \label{fig:cyb_attack}
    \vspace{-1mm}
\end{figure}
\vspace{-3mm}
\subsection{Mitigating Adversarial Workload Synchronization}
\vspace{-1mm}
In the first experiment, the \ac{dc} is placed at bus~374 (Riyadh New South Industrial, 380~kV), the location identified by~\eqref{eq:opt}, and its \ac{phil} response is shown in Fig.~\ref{fig:cyb_attack}. The power drawn by the 20 nodes is measured and scaled by a constant gain, such that the facility draws 1~GW on average. As a worst case, the adversary controls all 20 computation nodes and aligns their workloads: the facility demand then alternates between about 0.82 and 1.1~GW, while the transients of the server power supplies at each transition extend its swing to about 0.7 and 1.4~GW. The adversary, having briefly rented the workstations, conceals its intent by running \ac{llm} training at dispersed frequencies, whose oscillations dephase and aggregate into a nearly constant demand, as confirmed in hardware. At approximately 7.5~s, it aligns the workloads in frequency, at 0.8~Hz, and in phase, producing a large oscillation. The proposed detection flags the irregularity within the first cycle, and the mitigation restacks the workloads before the oscillation peaks, reducing the RMS oscillation of the facility demand from 199 to 31~MW (84\%) and that of the Riyadh City-14 frequency from 67 to 20~mHz (70\%). The experiment thus demonstrates that workload restacking can be performed accurately under realistic hardware conditions, substantially flattening the demand of the \ac{dc}.



%
\vspace{-4mm}
\subsection{Coordinating Distributed \ac{dc} Workloads}
\vspace{-1mm}
To demonstrate the importance of coordinated workload execution, the \ac{dc} located in Riyadh is relocated to Dammam, and a second \ac{dc} is placed in Jeddah, each dedicated to a single \ac{llm} training workload with different settings, e.g., duty ratio and frequency: the Dammam facility oscillates at 0.8~Hz with peaks of 1.4~GW, and the Jeddah facility at 0.85~Hz with peaks of 1.9~GW. Their superposition thus remains periodic, as shown in Sect.~III, but the two workloads drift in and out of phase, so that the aggregate power beats with a period of 20~s. Moreover, the power of each \ac{dc} reaches a given bus through the network and machine dynamics in between, which alter its magnitude and phase, so that the same profiles are felt {differently} across the network. The objective of Algorithm~\ref{alg:mitig} is therefore adapted to a set of target buses $\mathcal{M}$, here the 47 380~kV buses of Riyadh, which lies between the two \acp{dc} and hosts none, with $\chi_m$ denoting the voltage magnitude of bus~$m\in\mathcal{M}$. More generally, consider a set of facilities $\mathcal{D}$, each treated as a single workload, where facility~$d\in\mathcal{D}$ is connected at bus~$k_d$, has demand spectrum $\tilde{\mathcal{P}}_d$, and is shifted by a delay $\tau_d$. By superposition, the spectrum of $\chi_m$ is $\sum_{d\in\mathcal{D}}H_{k_d,m}(\mathrm{j}2\pi f)\,\tilde{\mathcal{P}}_d(f)\,e^{-\mathrm{j}2\pi f\tau_d}$, where $H_{k,m}$ is given by~\eqref{eq:tf}. Once all facilities are brought to a common base frequency $f_b$ or its harmonics, the objective becomes:
\begin{equation}\small
    \!\!\!\!J(\bfg\tau)\!\!=\!\!\!\!\!\sum_{m\in\mathcal{M}}\!\!\!\!w_m\!\!\sum_{h\geq1} \!\!w_h\Big|\!\sum_{d\in\mathcal{D}}\!\!H_{k_d,m}(\mathrm{j}h\omega_b)\,\tilde{\mathcal{P}}^{\prime}_d(hf_b)\,e^{-\mathrm{j}h\omega_b\tau_d}\!\Big|^2\!\!,
    \label{eq:split_changed}
\end{equation}
where $\omega_b=2\pi f_b$ and $w_m\geq0$ weights the importance of bus~$m$. The transfer functions thus enter only as fixed complex gains on the harmonics of each facility, so that $J$ retains the form of~\eqref{eq:mitig} and is minimized by the same algorithm.
In this study, $\mathcal{D}$ comprises the Dammam and Jeddah facilities, $f_b=0.8$~Hz, and all target buses are weighted equally, $w_m=1$.
\begin{figure}[t]
    \centering
    \includegraphics[width=0.8\columnwidth]{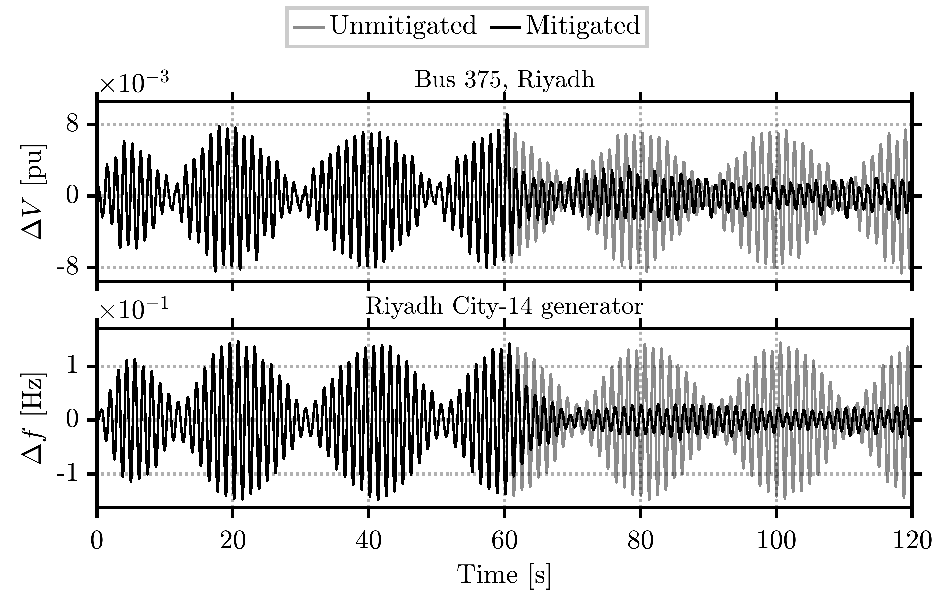}
    \vspace{-2mm}
    \caption{Voltage of bus~375 (top) and frequency of the Riyadh City-14 generator (bottom) under distributed \ac{dc} synchronization.}
    \label{fig:dc_work}
    \vspace{-1mm}
\end{figure}

The case study results are shown in Fig.~\ref{fig:dc_work} for the voltage of bus~375, Riyadh's largest load bus, and the frequency of the Riyadh City-14 generator, 43~km away. The mitigation plan is computed in 0.32~s and reduces the RMS voltage oscillation of the 47 target buses by 47-76\%, with a median of 61\%, and that of bus~375 by 68\%. The frequency oscillation of the City-14 generator, which is not part of the objective, falls by 77\%, since both \acp{dc} force the network at 0.8~Hz, to which this machine responds. This comes at the cost of slowing the Jeddah workload by 6\% to share the base frequency, retaining 96.8\% of the facility-power-weighted throughput.



%
\vspace{-3mm}
\section{Conclusions}
\vspace{-1mm}
This work showed that multi-tenant \acp{dc}, commonly regarded as benign, may oscillate even under normal operation, and an adversary renting their workstations can impose the same oscillation deliberately, at a cost that scales with its amplitude. Owing to their scale and persistence, such oscillations can interact with the controllers of grid assets and amplify their own impact, a risk that remains invisible as long as these facilities are assessed by their installed capacity rather than by the behavior of their workloads. Since the threat originates within the facility, it is there that it is most effectively contained: the proposed detection and mitigation mechanisms turn the same control an adversary exploits into a safeguard, trading throughput for a substantially smoother demand, and provide a foundation for policies governing workload execution. Treating workload execution as a grid-relevant activity is thus key to turning \acp{dc} from latent sources of oscillations into grid-aware infrastructure. 
\vspace{-4mm}
\bibliographystyle{IEEEtran}

\begingroup
\renewcommand{\baselinestretch}{0.95}\selectfont
\bibliography{references_compact}
\endgroup
\end{document}